\documentclass[aip,jcp,reprint]{revtex4-1}

\usepackage{graphicx}
\usepackage{bm}
\newcommand{\subscript}[1]{\ensuremath{_{\textrm{\footnotesize{#1}}}}}
\usepackage{amssymb}

\newcommand*{\ket}[1]{\left| #1 \right\rangle}

\newcommand*{\bra}[1]{\left\langle #1 \right|}

\begin{document}
\title{Calculations of Potential Energy Surfaces Using Monte Carlo Configuration Interaction}

\author{Jeremy P. Coe}
\author{Daniel J. Taylor}
\author{Martin J. Paterson}
\email{m.j.paterson@hw.ac.uk}
\affiliation{ 
Institute of Chemical Sciences, School of Engineering and Physical Sciences, Heriot-Watt University, Edinburgh, EH14 4AS, UK.
}%

\begin{abstract}
We apply the method of Monte Carlo configuration interaction (MCCI) to calculate ground-state potential energy curves for a range of small molecules and compare the results with full configuration interaction.  We show that the MCCI potential energy curve can be calculated to relatively good accuracy, as quantified using the non-parallelity error, using only a very small fraction of the FCI space.  In most cases the potential curve is of better accuracy than its constituent single-point energies.  We finally test the MCCI program on systems with basis sets beyond full configuration interaction: a lattice of fifty hydrogen atoms and ethylene.  The results for ethylene agree fairly well with other computational work while for the lattice of fifty hydrogens we find that the fraction of the full configuration interaction space we were able to consider appears to be too small as, although some qualitative features are reproduced, the potential curve is less accurate.
\end{abstract}

\maketitle

\section{Introduction}
Full configuration interaction (FCI) enables, in theory, a quantum system to be modelled as accurately as possible within a given basis set, but the rapidly increasing number of configurations means that such calculations are out of reach except for sufficiently small molecules and basis sets.  Standard truncation methods, such as considering only single and double excitations with respect to a reference state or restricting excitations to a pre-defined space, can reduce the size of the calculation but often at the expense of the accuracy and consistency of the correlation energy as important configurations may be neglected.  

  It is acknowledged, however, that a large proportion of the states in many full configuration interaction wavefunctions tend to have practically negligible coefficients and should not be expected to individually contribute substantially to the properties of the wavefunction. Novel approaches to truncated configuration interaction have been developed to attempt to tackle the difficulties posed by the exceedingly rapid growth of the standard full CI space by exploiting this observation and trying to seek out the important states, see Ref.~\onlinecite{NovelTruncatedCI} for a recent review.  For example these include a priori estimates,\cite{aprioriRuedenberg,aprioriFranceshetti} Monte Carlo sampling of a DMRG calculation\cite{DMRGestReiher11} and estimates from perturbation theory.\cite{HarrisonFCIperturbation}  Another promising method is that of Monte Carlo configuration interaction (MCCI) developed originally by Greer.\cite{MCCIGreer95,MCCIcodeGreer}   In MCCI a trial wavefunction is randomly augmented with coupled states in an iterative scheme where those states with coefficients smaller than a certain value in the resulting solution of the Schr\"{o}dinger equation are eventually removed.  Even without prior knowledge of the important configurations or molecular orbitals such a procedure can, in principle, result in a compact wavefunction which recreates much of the energy of the FCI wavefunction but using only a small fraction of the FCI states.  

Single-reference methods based on coupled-cluster\cite{ShavittBartlettCCBook,*CrawfordShaeferCC} such as coupled-cluster singles doubles (CCSD)\cite{CCSD} are considered to have some of the best efficiencies in calculations for systems in which the FCI expansion would have one dominant configuration and the correlation is considered dynamic. However, they may have problems---especially if the restricted Hartree-Fock determinant is used as a reference---when there are a number of important configurations and the system is deemed multireference. These configurations are associated with static correlation and, e.g., dissociation energies may be modelled incorrectly.  For multireference systems the associated static correlation may be accounted for by using methods such as complete active space self-consistent field (CASSCF),\cite{siegbahn:2384} but insight is required into the selection of orbitals for the active space and the calculation becomes intractable as the size of the active space increases. If the orbitals have been selected appropriately then perturbation methods (CASPT2) or multi-reference configuration interaction (MRCI) may then be used to account for the remaining dynamic correlation but at relatively high cost.  In principle MCCI can give a compact wavefunction which can account for both static and dynamic correlation with minimal user choices: the accuracy and size of the calculation is essentially controlled only by the coefficient cut-off parameter $c_{\text{min}}$.

MCCI has been successfully applied to the single point energy of the water molecule,\cite{MCCIGreer95} and the dissociation energy of HF and H\subscript{2}O.\cite{dissociationGreer}   Electronic excitation energies have also been calculated\cite{excite1Greer} for the first-row atoms beginning with carbon---and also for silicon---with relatively small average errors and numbers of states.  While the excitation energies of molecules CH\subscript{2}, C\subscript{2}, N\subscript{2} and H\subscript{2}O  were considered in Ref.~\onlinecite{GreerMCCISpectra} with only tens of meV error and using only a very small proportion of the states compared with FCI.

In this work we investigate the ability of MCCI to calculate potential energy curves.  If the MCCI results can have an almost balanced error across different geometries then we should be able to produce a curve of better accuracy, as quantified by the non-parallelity error, than its constituent single-point energies.  We first assess the usefulness of this method on  molecules with basis sets for which FCI results exist. We look at potential curves with respect to the dissociation of one hydrogen from HF, BH, and CH\subscript{4}. The curves for bond stretches are produced for C\subscript{2}, F\subscript{2}, N\subscript{2} and H\subscript{12}. The potential curves for NH\subscript{3} inversion and the model formation of BeH\subscript{2} along the reaction coordinate are also calculated. Finally systems beyond the scope of current FCI are also investigated.  We calculate a potential curve for H\subscript{50} which we compare with density-matrix renormalization group (DMRG)\cite{DMRGchemWhite99} results, and a potential curve for ethylene as the torsional angle is varied which we compare with other computational results.   

\section{Method}

A FCI wavefunction may be written in the notation of second quantization as
\begin{eqnarray}
\nonumber \ket{\Psi}=c_0 \ket{\Psi_0} +\sum_{i,j} c_i^{j} a_{j}^{\dagger}a_{i} \ket{\Psi_0}\\+
\sum_{k<i,l<j} c_{ik}^{jl}  a_{l}^{\dagger}a_{k}a_{j}^{\dagger}a_{i} \ket{\Psi_0} +\cdots
\label{eqn:FCIwave}
\end{eqnarray}
where  $a_{i}^{\dagger}$ ($a_{i}$) creates (annihilates) the orbital $i$ in a state. Here letters appearing as subscripts on the coefficients label occupied orbitals in the reference state $\ket{\Psi_{0}}$  while superscripts are unoccupied orbitals in the reference. A traditional truncation would limit the calculation size by only including substitutions up to a certain substitution level. Limiting the expansion to only the depicted terms in Eq.\ref{eqn:FCIwave} would correspond to a CISD calculation.  MCCI instead randomly augments a wavefunction $\ket{\Psi}=\sum_{i} c_{i} \ket{\psi_{i}}$ with coupled configurations, i.e., states $\bra{\psi_{j}}$ such that $\bra{\psi_{j}} \hat{H} \ket{\Psi} \neq 0$ so that, in principle, the important configurations can eventually be found regardless of their substitution level. Configuration state functions (CSFs) are used in MCCI rather than the more usual Slater determinants (SDs).  This means that the MCCI wavefunction fulfils at least one of the requirements of the exact wavefunction: it is an eigenfunction of $\hat{S}^{2}$.  In addition fewer states are needed, e.g., the ratio of SDs to the number of CSFs\cite{PaldusCSFform} when $S=M_{s}=0$ for $8$ electrons and $30$ basis functions is around $4.4$. One downside is that the calculation of the overlap and Hamiltonian matrix is more complicated than when using SDs.\cite{Harris74} 

  We briefly describe the full procedure for the MCCI algorithm\cite{MCCIGreer95,MCCIcodeGreer} which consists of branching, diagonalization and pruning.  In the branching step, coupled CSFs, created by a symmetry preserving stochastic single or double substitution with equal probability in the current set of CSFs, are added. Branching is always attempted from CSFs with a coefficient greater than a certain value while other CSFs each have probability one half of being used.  The coefficients are then found for the approximate wavefunction expanded in this augmented set of CSFs by solving the generalised eigenvalue problem of $H\bm{c}=ES\bm{c}$.  Here $H$ and $S$ are the Hamiltonian and overlap matrix within this subspace.  Finally there is a pruning step: added CSFs with coefficients less than a certain value, $c_{\text{min}}$, in the wavefunction expansion are then discarded.  These three steps are repeated for a large enough number of iterations so that the energy appears to have converged.  In addition, every $k$ iterations (where $k=10$ in this work) a pruning step is implemented whereby all of the CSFs in the wavefunction are also considered as candidates for removal depending on the magnitude of their coefficient.  Furthermore there is no branching or pruning on the last iteration, but a full pruning step is carried out on the penultimate iteration. The program can run in parallel where states that have been added and retained at each step on each processor are shared with all other processors.  The results in this work come from MCCI calculations using either eight or twelve processors.

We initialise the procedure with the CSF formed from the restricted Hartree-Fock wavefunction within a given basis.  New Hartree-Fock orbitals are calculated at each geometry.  For the calculation of the Hartree-Fock molecular orbitals and their one-electron and two-electron integrals we use the MOLPRO package.\cite{MOLPRO}

\section{Results}

As the potential is defined up to an additive constant then it is the shape of the potential energy curve that is important. So two curves that differed only by a constant would for all practical purposes be the same. A useful approach to quantify the accuracy of the MCCI potential curve is therefore the non-parallelity error (NPE) \cite{li:1024NPE} defined here for an approximate energy $E^{\text{approx}}$ as 
\begin{equation}
NPE=\max_{R} |E^{\text{FCI}}_{R}-E^{\text{approx}}_{R}|-\min_{R} |E^{\text{FCI}}_{R}-E^{\text{approx}}_{R}|
\end{equation}
where $R$ ranges over all considered bond lengths.

Hence if we can achieve a balanced accuracy in the energy across the geometries we can recover a potential energy curve with higher accuracy than its single-point energies.  We note that this is not necessarily guaranteed as if many points are essentially the FCI energy but there are a few points with large errors then the NPE would be high but the mean single-point energy error could be relatively low.  To investigate this we attempt to use a small enough $c_{\text{min}}$, where computationally feasible, to produce a sufficiently accurate curve.

\subsection{Hydrogen dissociation}

We first consider varying the length of a bond with hydrogen in BH, HF, and CH\subscript{4} and compare the potential energy curves with FCI results from Ref.~\onlinecite{SherrillHbondFCI}. These potential curves represent one of the simplest dissociations: the breaking of a bond to hydrogen.  As such they would be expected to have some multireference character and make an interesting test case for MCCI to be compared with other methods.  For these three systems one core orbital is frozen in all the results.

\subsubsection{HF}

\begin{figure}[ht]\centering
\includegraphics[width=.45\textwidth]{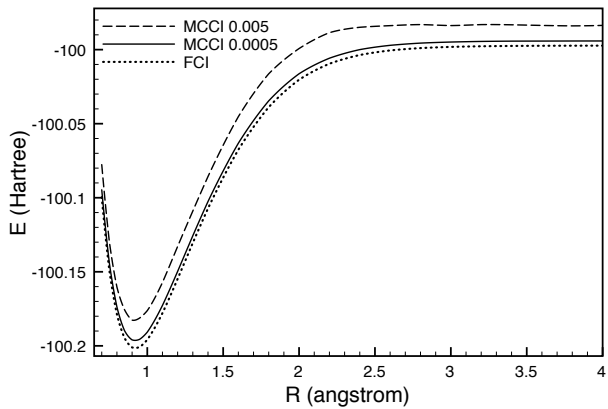}
\caption{MCCI and FCI results\cite{SherrillHbondFCI} for energy (Hartree) against bond length $R$ (angstrom) for HF with the 6-31G** basis set and one frozen core orbital.}\label{fig:HFpot}
\end{figure}

We see in Fig.~\ref{fig:HFpot} that the shape of the HF FCI curve in a 6-31G* basis is reproduced by MCCI and it appears that for this system, apart from the energy shift, there is not much difference between the two cut-off values.  The reduction in NPE by decreasing the cut-off value from $c_{\text{min}}=5\times10^{-3}$ to $c_{\text{min}}=5\times10^{-4}$ is $5.7$ to $1.3$ kcal/mol.  These NPE compare very favourably with methods based on a single reference where CCSD gives a value of $13.0$ kcal/mol, the CCSD(T) NPE is as high as $26.8$ kcal/mol, while that of UCCSD(T) is $3.7$ kcal/mol.\cite{SherrillHbondFCI}  For the multireference approaches considered in Ref.~\onlinecite{AbramsMulti03}, an NPE of $17.69$ kcal/mol was found for CASSCF(8, 5) `valence active space' and $4.83$ kcal/mol for CASSCF(8, 8) `1:1 active space'. CASPT2(8, 5) and CASPT2(8, 8) resulted in NPEs of $2.77$ and $0.5$ kcal/mol. The NPEs for SOCI(8, 5) and SOCI(8, 8) were $3.20$ and $0.04$ kcal/mol respectively. Ref.~\onlinecite{SearsMulti05} looked into using a minimal active space (one bonding orbital and one anti-bonding orbital for each broken bond) and although their CASSCF(2, 2) had a higher NPE (18.66 kcal/mol) than when using the aforementioned active spaces, the CASPT2 value was lower at $0.47$ kcal/mol and the SOCI was better than when using the larger valence active space but, at $0.98$ kcal/mol, not as good as when using the 1:1 active space.

  The larger cut-off for MCCI required $172.3$ CSFs on average, while the smaller needed $1,337$.  The full configuration space consisted of $3,756,816$ SDs\cite{SherrillHbondFCI} so only a very small fraction were used by MCCI to give an NPE of a few kcal/mol (see table~\ref{tbl:CIsizeHdiss}).

\begin{figure}[ht]\centering
\includegraphics[width=.45\textwidth]{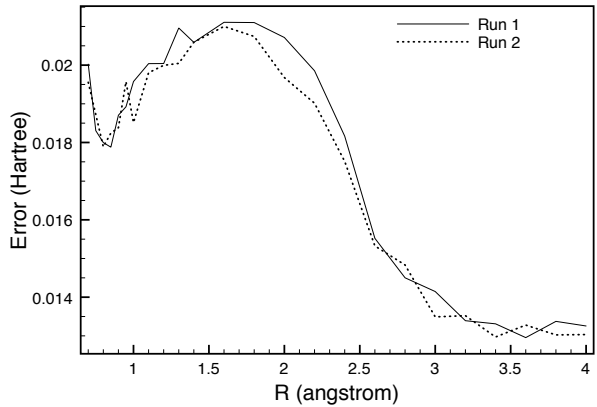}
\caption{Error in two MCCI runs with $c_{\text{min}}=5\times10^{-3}$ compared with the FCI results\cite{SherrillHbondFCI} against bond length $R$ (angstrom) for HF with the 6-31G** basis set and one frozen core orbital.}\label{fig:HFpoterror1}
\end{figure}

\begin{figure}[ht]\centering
\includegraphics[width=.45\textwidth]{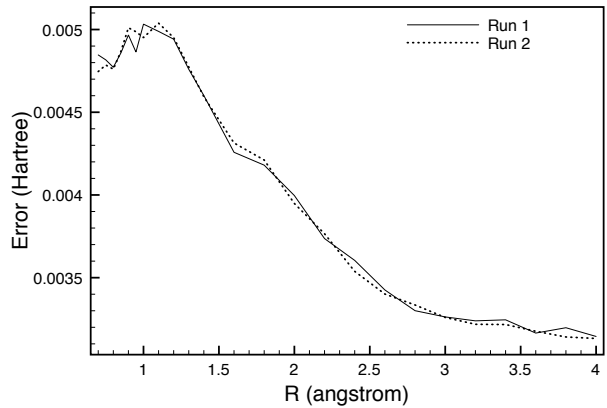}
\caption{Error in two MCCI runs $c_{\text{min}}=5\times10^{-4}$ compared with the FCI results\cite{SherrillHbondFCI} against bond length $R$ (angstrom) for HF with the 6-31G** basis set and one frozen core orbital.}\label{fig:HFpoterror2}
\end{figure}

 Figures~\ref{fig:HFpoterror1} and \ref{fig:HFpoterror2} show that the MCCI results are closest to the FCI when the bond length is large.  We also see that rerunning the calculations with a different random number seed does not change the error much and the variation in error seems to be lower as the cut-off is decreased.

As the MCCI potential curves (Fig.~\ref{fig:HFpot}) appear to have essentially reached dissociation, we may compare MCCI energies at $R=4$ angstrom with the total energy from appropriate MCCI calculations on the fragments to indicate how close the MCCI calculation is to size consistency.  We find a difference of $4.0$ kcal/mol for the larger cut-off and  $0.15$ kcal/mol for the smaller cut-off. 
\subsubsection{BH}

\begin{figure}[ht]\centering
\includegraphics[width=.45\textwidth]{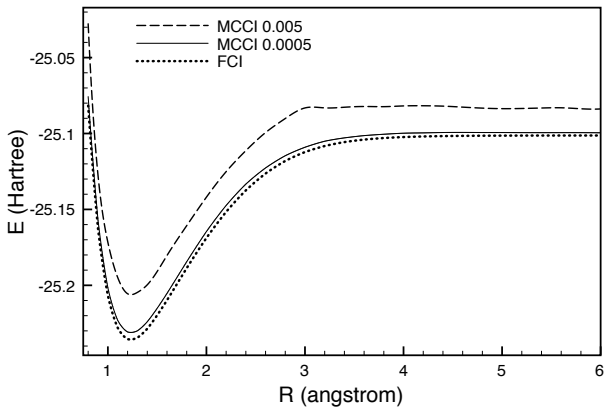}
\caption{MCCI and FCI results\cite{SherrillHbondFCI} for energy (Hartree) against bond length $R$ (angstrom) for BH with the aug-cc-pVQZ basis set and one frozen core orbital.  Adapted from Ref.~\onlinecite{NovelTruncatedCI}.}\label{fig:BHpot}
\end{figure}

The results for BH were calculated in an aug-cc-pVQZ basis.  Preliminary MCCI results for BH are contained in Ref.~\onlinecite{NovelTruncatedCI}. In Fig.~\ref{fig:BHpot} we can see that with a cut-off of $5\times10^{-3}$ the general shape of the MCCI potential curve is qualitatively correct with the minimum in approximately the right place and a well-behaved dissociation although there is a small shoulder in the potential around $R=3$ angstrom.  The limiting value of the energy is reached too early however. By decreasing $c_{\text{min}}$ to $5\times10^{-4}$ the curve is almost that of the FCI shifted by a small energy.  The NPE is $22.8$ kcal/mol in the former case and reduces to $2.6$ kcal/mol with the smaller $c_{\text{min}}$ value.  For this value of $c_{\text{min}}$ the mean number of CSFs was only $4,220$ compared with around fifteen million Slater determinants in the FCI space (See Table~\ref{tbl:CIsizeHdiss}) while $330$ were required on average for the larger case.  From Ref.~\onlinecite{SherrillHbondFCI} the NPE value for CCSD was $8.1$ kcal/mol, that of CCSD(T) was $23.3$ kcal/mol while UCCSD(T) had an NPE as low as $3.1$ kcal/mol.  Multireference methods have also been considered\cite{AbramsMulti03} and compared with FCI results for the potential curve of BH but with a smaller range and a cc-pVQZ basis as there were difficulties in convergence when using the aug-cc-pVQZ basis.  Although not a direct comparison, we note that the CASSCF(4, 4) `valence active space' there had an NPE of $12.68$ kcal/mol, CASSCF(4, 5) `1:1 active space' gave $9.38$ kcal/mol, CASPT2(4, 5) gave $3.16$  kcal/mol while only the second-order CI (SOCI) results were lower than the MCCI value in this work at $0.29$ and $1.54$ kcal/mol for active spaces of (4, 4) and (4, 5) respectively.

\begin{figure}[ht]\centering
\includegraphics[width=.45\textwidth]{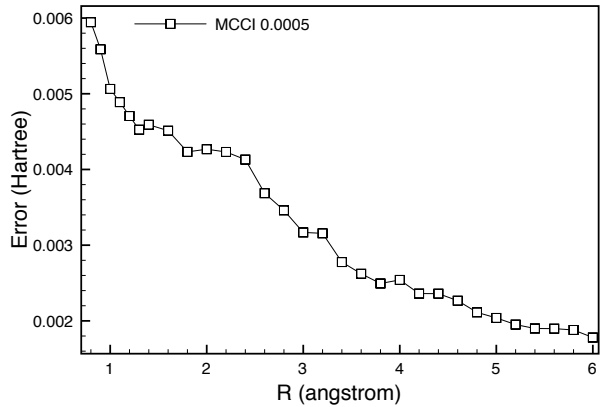}
\caption{MCCI ($c_{\text{min}}=5\times10^{-4}$) energy error (Hartree) when compared with FCI results against the bond length $R$ (angstrom) for BH with the aug-cc-pVQZ basis set and one frozen core orbital. The NPE is $2.6$ kcal/mol.}\label{fig:BHerror}
\end{figure}

We see in Fig.~\ref{fig:BHerror} that the energy error for this system at $c_{\text{min}}=5\times10^{-4}$ when compared with FCI generally decreases as the bond length increases. We note that the difference between the MCCI energy at $R=6$ angstrom and the total MCCI energy for the fragments is $6.6$ kcal/mol at $c_{\text{min}}=5\times10^{-3}$ and $0.83$ kcal/mol at $c_{\text{min}}=5\times10^{-4}$.

\subsubsection{CH\subscript{4}}
\begin{figure}[ht]\centering
\includegraphics[width=.45\textwidth]{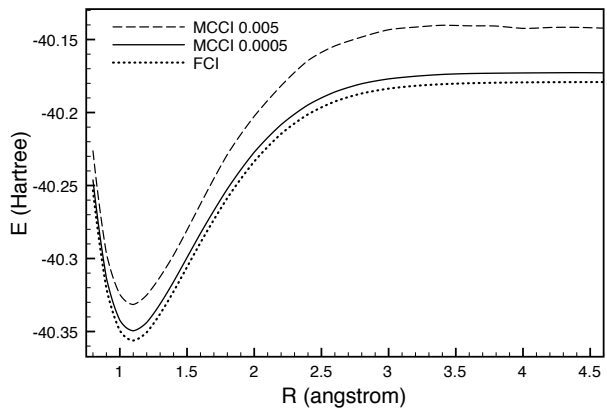}
\caption{MCCI and FCI results\cite{SherrillHbondFCI} for energy (Hartree) against one carbon-hydrogen bond length $R$ (angstrom) for CH\subscript{4} with the 6-31G* basis set and one frozen core orbital. Adapted from Ref.~\onlinecite{NovelTruncatedCI}.}\label{fig:CH4pot}
\end{figure}

As in Ref.~\onlinecite{SherrillHbondFCI} a tetrahedral geometry was used for methane with bond length $1.086$ angstrom for the three carbon-hydrogen bonds that were not varied.  Preliminary MCCI results for CH\subscript{4} are contained in Ref.~\onlinecite{NovelTruncatedCI}.  Fig.~\ref{fig:CH4pot} shows that, in a 6-31G* basis, MCCI again captures the shape of the FCI potential curve. Now it appears that the larger cut-off may be a little too high in energy at large $R$ compared with the equilibrium geometry.  This is quantified using the NPE which is $10.3$ kcal/mol for $c_{\text{min}}=5\times10^{-3}$ but drops to $0.6$ kcal/mol at the smaller cut-off.  We note that in Ref.~\onlinecite{SherrillHbondFCI}  CCSD gave a value of $10.3$ kcal/mol, UCCSD gave $5.1$ kcal/mol while UCCSD(T) was as low as $3.2$ kcal/mol. In Ref.~\onlinecite{AbramsMulti03} the definition of the 1:1 active space was equivalent to the valence active space in this case and gave CASSCF(8, 8) with an NPE of $6.34$ kcal/mol, CASPT2(8, 8) gave $1.56$ kcal/mol,  CISD[TQ](8, 8) gave $1.33$ kcal/mol, while only SOCI(8, 8) had a lower NPE than the MCCI results with $0.3$ kcal/mol. The minimal active space results\cite{SearsMulti05} had an NPE of $9.25$ kcal/mol for CASSCF(2,2), with CASPT2 at $1.22$ kcal/mol and SOCI a little worse with $0.6$ kcal/mol. Hence only one of the results from these two works using multireference methods gave a lower NPE for methane.

 The FCI space is around $26.7$ million SDs,\cite{SherrillHbondFCI} but in Table~\ref{tbl:CIsizeHdiss} we see that the MCCI results are achieved using only $417$ and $4,272$ CSFs on average for the larger and smaller $c_{\text{min}}$ respectively. 

\begin{figure}[ht]\centering
\includegraphics[width=.45\textwidth]{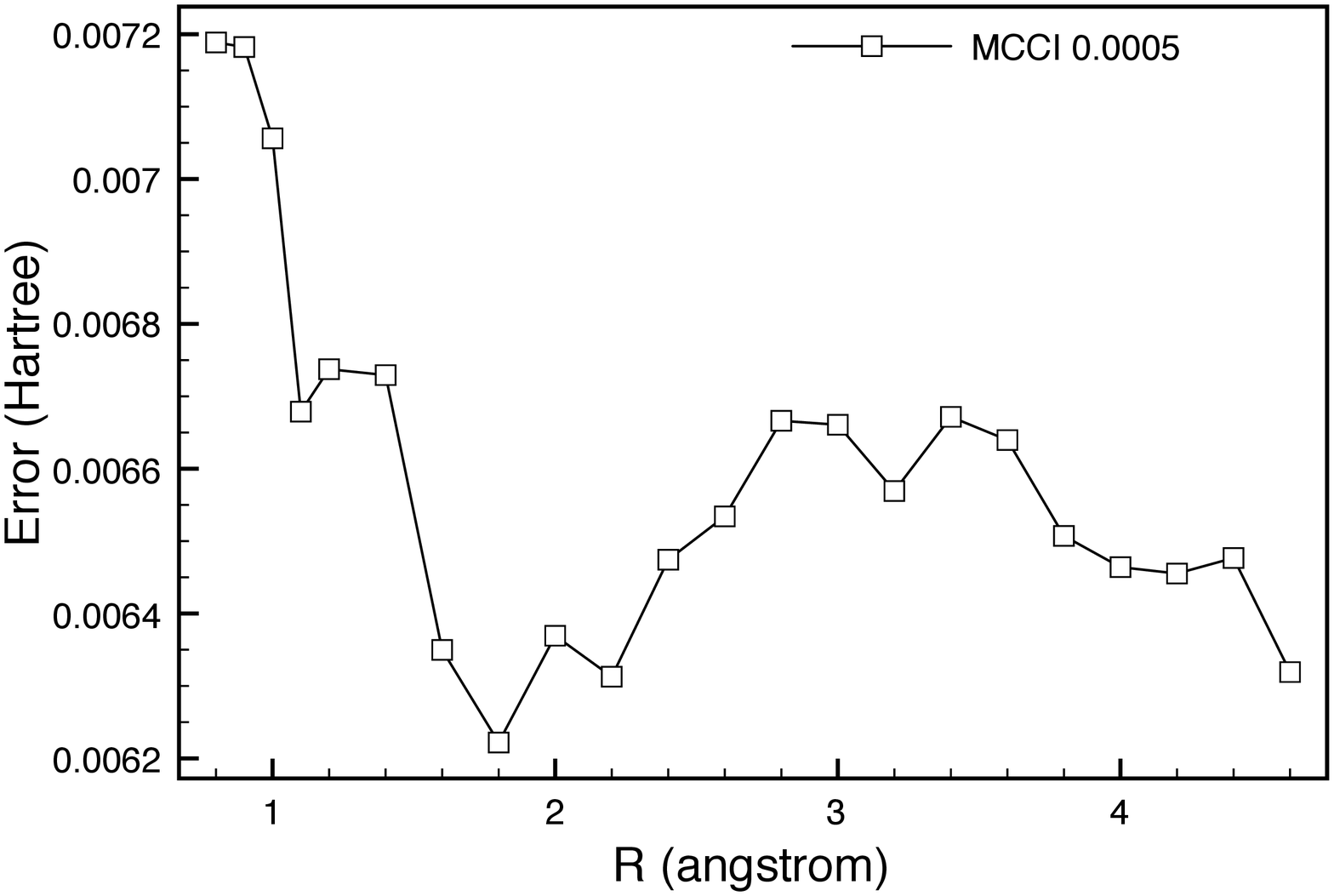}
\caption{MCCI ($c_{\text{min}}=5\times10^{-4}$) energy error (Hartree) when compared with FCI results against one carbon-hydrogen bond length $R$ (angstrom) for CH\subscript{4} with the 6-31G* basis set and one frozen core orbital.  The NPE is $0.6$ kcal/mol.}\label{fig:CH4error}
\end{figure}

In Fig.~\ref{fig:CH4error} we see that the error in the energy compared with FCI is greatest at very small bond lengths and now the minimum error is at bond lengths a little longer than that at equilibrium, but the errors are all very small.  The difference in energies between MCCI applied to the system at $R=4.6$ angstroms and the sum of MCCI applied to the fragments is 15 kcal/mol at the larger cut-off and 0.79 kcal/mol at the smaller cut-off.  For HF, BH and CH\subscript{4}  at $c_{\text{min}}=5\times10^{-3}$ the difference between the FCI and MCCI energy, at all geometries considered, ranged from around $8.4$ kcal/mol to $33.8$ kcal/mol while this range was around $1.1$ kcal/mol to $4.5$ kcal/mol when the cutoff was lowered to $c_{\text{min}}=5\times10^{-4}$. 

\begin{table}[h]
\centering
\caption{MCCI mean CSFs compared with FCI symmetry-adapted SDs} \label{tbl:CIsizeHdiss}
\begin{tabular*}{9.5cm}{@{\extracolsep{\fill}}lccc}
\hline
\hline
Method &  BH states & HF states  & CH\subscript{4} states   \\
\hline
  MCCI0.005 (CSFs) & 333.1 &  172.3 & 417.0  \\
 MCCI0.0005 (CSFs)  & 4219.6 & 1337.0 & 4272.0\\
  FCI (SDs) &$15,132,412$ & $ 3,756,816 $ & $26,755,625$ \\
   MCCI0.005 fraction & $0.0022\%$ &$0.0046\%$ &$0.0016\%$\\
   MCCI0.0005 fraction & $0.028 \%$ & $0.036\%$& $0.016\%$ \\
\hline
\hline
\end{tabular*}
\end{table}

We have seen, for the three systems considered for hydrogen dissociation, that by going to a low enough coefficient cut-off value of $c_{\text{min}}=5\times10^{-4}$ we can reproduce the shape of the potential curve to relatively high accuracy: we achieve a non-parallelity error of around a few kcal/mol which is  better than the  single-reference methods we compare with from the literature.   This is accomplished using in the order of hundredths of a percent of the full CI space.  The accuracy was also higher than many of the multireference results from the literature.  Some CASPT2 and SOCI calculations, using large active spaces or chemical intuition in the construction of a smaller active space, produced a lower non-parallelity error but the number of states used for these calculations was not reported for comparison.

\subsection{Carbon Dimer}

\begin{figure}[ht]\centering
\includegraphics[width=.45\textwidth]{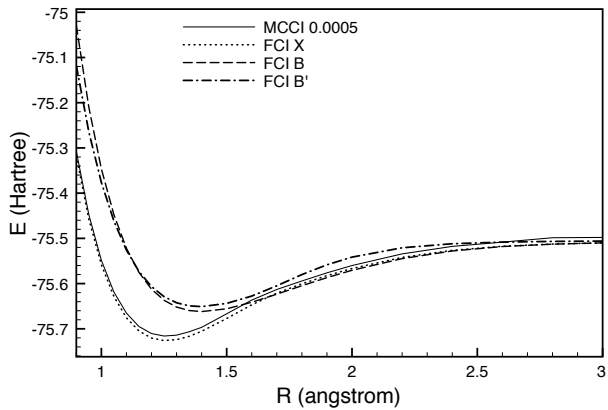}
\caption{MCCI ($c_{\text{min}}=5\times10^{-4}$) and FCI\cite{SherrillC2FCI} results for energy (Hartree) against the bond length $R$ (angstrom) for C\subscript{2} using the 6-31G* basis set with two frozen core orbitals.  Here X and B' are $^{1}\Sigma_{g}^{+}$ and B is $^{1}\Delta_{g}$.  Adapted from Ref.~\onlinecite{NovelTruncatedCI}.}\label{fig:C2pot}
\end{figure}

Next we look at the dissociation of C\subscript{2} in a 6-31G* basis with two frozen orbitals.  This system is known to be multireference and possess low-lying excited states hence poses a stern test of MCCI.  Preliminary MCCI results for C\subscript{2} are contained in Ref.~\onlinecite{NovelTruncatedCI}.  FCI results\cite{SherrillC2FCI} and the MCCI potential curve with $c_{\text{min}}=5\times 10^{-3}$ are shown in Fig.~\ref{fig:C2pot}.  The MCCI curve accuracy is somewhat difficult to judge by eye due to the many close excited states, but it appears to follow the FCI ground state closely except for bond lengths ($R$) around $3$ angstrom where it seems that a low lying excited state (B') has been converged upon instead.  However it appears that the three FCI curves are in fact tending to a degenerate state as the system dissociates.  In addition, the lowest energy state between about $1.7$ and $2.5$ angstrom is B rather than X and MCCI may have converged to this as would be expected: the states have different symmetry so can cross but have the same spatial symmetry in the abelian subgroup ($D_{2h}$) used for computation.

The mean number of CSFs employed was $\sim 6,900$ for MCCI here, while  FCI results from Ref.~\onlinecite{SherrillC2FCI} required $52,407,353$ symmetry-adapted Slater determinants. This gives a ratio of $0.013\%$. The NPE is $4.9$ kcal/mol for MCCI here which is substantially better than methods based on a single reference. From Ref.~\onlinecite{SherrillC2FCI} CCSD had an NPE of $24.3$ kcal/mol and that of CCSD(T) was $61.3$ kcal/mol, highlighting the instability of coupled cluster at large $R$. Unrestricted methods behaved better at dissociation but were less good at intermediate bond lengths resulting in a NPE for UCCSD(T) of $21.6$ kcal/mol. CISDTQ gave a curve with $16.6$ kcal/mol for the NPE.  It is worth noting that a CISDTQ, without symmetry, and using Slater determinants would be expected to have around three million terms. Even allowing for the reduction in the size of the space through spatial symmetry and the use of CSFs, the MCCI result with circa $6,900$ CSFs on average and a better NPE shows the usefulness of the MCCI approach here.  Valence active space CASSCF in Ref.~\onlinecite{SherrillFCIandmulti} used $660$ determinants and had an NPE of $5.4$ kcal/mol for (SA)-CASSCF. (EOM)CCSD gave $24.3$ kcal/mol and CR-(EOM)CCSD(T),III gave $13.5$ kcal/mol for the NPE.  CISD[TQ] had an NPE of $16$ kcal/mol in that work and used $87,415$ determinants. In fact only two of the multireference techniques investigated in Ref.~\onlinecite{SherrillFCIandmulti} had a lower NPE: (SA)-CASSCF MRCI with $0.4$ kcal/mol using $270,338$ determinants and SA-CASPT2 with an NPE of $3.8$ kcal/mol.    

A wavefunction for the carbon dimer found using FCI quantum Monte Carlo (FCIQMC) in Slater determinant space in Ref.~\onlinecite{AlaviFCIQMCcarbondimer} required $9.15 \times 10^{6}$ walkers at $0.9$ angstrom in this 6-31G* basis to give almost FCI results although the point at $R=3$ angstrom was neglected.  We note that with a cc-pVDZ basis and requiring that only states with zero angular momentum were allowed they calculated very similar results to MRCI, and with lower energy for $R>1.15$ angstrom than when using 6-31G*, with $2\times 10^{6}$ walkers compared with a standard FCI space of $4.74 \times 10^{9}$ Slater determinants.  This suggests that the performance of MCCI here could perhaps be improved if we also discriminate between states using their angular momentum in addition to their symmetry and spin.

\begin{figure}[ht]\centering
\includegraphics[width=.45\textwidth]{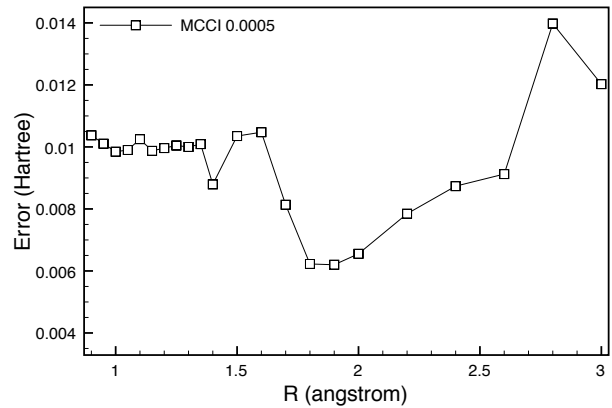}
\caption{MCCI ($c_{\text{min}}=5\times10^{-4}$) energy error (Hartree) when compared with FCI\cite{SherrillC2FCI} results against the bond length $R$ (angstrom) for C\subscript{2} using the 6-31G* basis set with two frozen core orbitals. The NPE is $4.9$ kcal/mol.}\label{fig:C2error}
\end{figure}

We note that MCCI potential curve for the carbon dimer is of slightly higher accuracy, in a sense, than its constituent single-point energy calculations as the mean single-point error was around $6.0$ kcal/mol.  In addition, if we exclude the two points around $R=3$, that seem to have converged to a low-lying excited state, then the single point energy error only reduces a little to $5.7$ kcal/mol but the NPE is now just $2.7$ kcal/mol.  In Fig.~\ref{fig:C2error} we see that the error does not vary too much, as summarised by the NPE, but is highest approaching the largest bond length.  Interestingly it is lowest not at equilibrium but at bond lengths from around $1.8$ to $2$ angstroms, which in this case may be due to the lowest energy state being reached, not the original ground-state. This slightly lowers the mean single-point error in this case, and will slightly raise the NPE. At $3$ angstrom we compare the MCCI energy for the full system with that of the sum of the MCCI energies of the fragments and find a difference of $3.2$ kcal/mol. We note that as the energy of the system using MCCI is higher than the energy of the fragments here then as the system has perhaps not sufficiently dissociated the difference is perhaps smaller than if a larger $R$ had been considered.

\subsubsection{Analysis of the C\subscript{2} MCCI wavefunction}

At the equilibrium geometry for C\subscript{2} the MCCI wavefunction has states with as many as $7$ substitutions from the Hartree-Fock reference and the multiconfigurational nature is evident from the largest ten coefficients of the CSFs in Table~\ref{tbl:coeffsC2}.  By $R=2.0$ the dominant configuration is now a double substitution of the Hartree-Fock reference which could suggest that the lowest lying state here (B), not the equilibrium ground-state curve (X), may have been converged upon as expected or could just be an aspect of the multiconfigurational nature which is is more apparent here: the largest two coefficients are $-0.636$ and $0.557$.   By $R=3.0$ the largest coefficient belongs to a CSF which differs to the major contributor in the previous two cases.  This may be linked to the suggestion from the potential curves that here a different excited state is converged upon at large $R$.

\begin{table}[h]
\centering
\caption{Ten largest coefficients of the C\subscript{2} MCCI wavefunction at $R=1.25$ angstrom with the number of substitutions with respect to the Hartree-Fock reference listed by spin.  } \label{tbl:coeffsC2}
\begin{tabular*}{8.5cm}{@{\extracolsep{\fill}}lcc}
\hline
\hline
Coefficient&  $\alpha$ substitutions  & $\beta$ substitutions    \\
\hline
 -0.830 &               0  &         0 \\   
  0.331     &           1    &       1 \\  
 -0.184        &      1     &      1   \\
 -0.180       &       1     &      1  \\
 -0.179       &        1     &      1 \\
 -0.160      &         0     &      2 \\
  0.158     &          1     &      1 \\
  0.158   &           1     &      1 \\
 -0.142     &           0     &      1   \\ 
 0.106    &            2     &      2 \\
\hline
\hline
\end{tabular*}
\end{table}

\subsection{F\subscript{2}}
\begin{figure}[ht]\centering
\includegraphics[width=.45\textwidth]{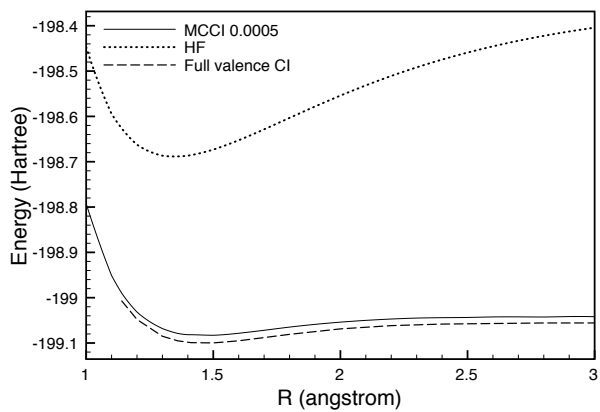}
\caption{MCCI (two frozen orbitals) with $c_{\text{min}}=5\times10^{-4}$, Full valence CI\cite{BytautasF2} and Hartree-Fock results for energy (Hartree) against bond length $R$ (angstrom) for F\subscript{2} using the cc-pVDZ basis set.}\label{fig:F2pot}
\end{figure}

In Fig.~\ref{fig:F2pot} we now display the MCCI potential curve for F\subscript{2} in a cc-pVDZ basis set without any frozen orbitals.  The MCCI results can be seen to follow the shape of the full valence CI results of Ref.~\onlinecite{BytautasF2} despite the MCCI calculation not freezing any orbitals.  We can also see how the Hartree-Fock energy increases much too rapidly as the bond length becomes large. 

 We note that the MCCI with $c_{\text{min}}=5\times 10^{-4}$ required around $3,577$ CSFs on average for the depicted points. The FCI space with two frozen orbitals would be expected to comprise $4.3\times 10^{11}$. Even though only a very small fraction of this space is used ($8.3\times 10^{-7}$), when compared with the Full valence CI points the NPE for MCCI is $6.2$ kcal/mol. 

\begin{figure}[ht]\centering
\includegraphics[width=.45\textwidth]{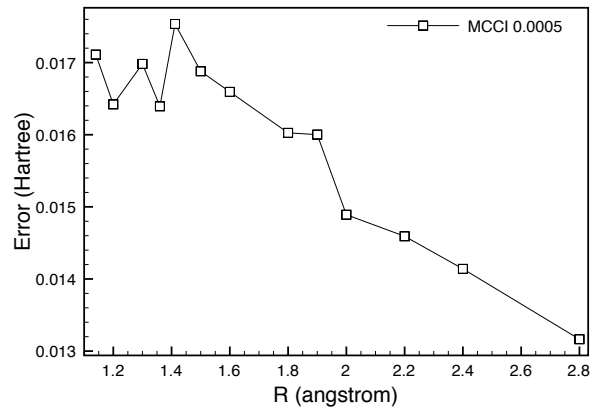}
\caption{Energy error of MCCI (two frozen orbitals) with $c_{\text{min}}=5\times10^{-4}$ when compared with Full valence CI\cite{BytautasF2} against bond length $R$ (angstrom) for F\subscript{2} using the cc-pVDZ basis set.  The NPE is $6.2$ kcal/mol.}\label{fig:F2poterror}
\end{figure}

In Fig.~\ref{fig:F2poterror} we display the difference between the MCCI and Full valence CI for the latter's points in the range displayed in Fig.~\ref{fig:F2pot}. Here the MCCI error tends to decrease with increasing bond length for the points displayed.  Comparing the MCCI energy of the system at $R=3$ angstrom with the sum of the MCCI energies of the constituent atoms we find a difference of $7.3$ kcal/mol.

\subsection{N\subscript{2}}

\begin{figure}[ht]\centering
\includegraphics[width=.45\textwidth]{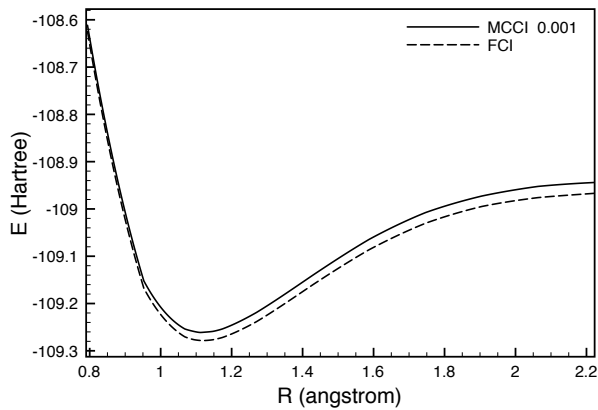}
\caption{MCCI ($c_{\text{min}}=10^{-3}$) and FCI \cite{LarsenN2FCI2000,GwaltneyN2FCI2002,chanN2FCI2002} energy (Hartree) against bond length $R$ (angstrom) for N\subscript{2} using the cc-pVDZ basis set with two frozen cores. }\label{fig:N2pot}
\end{figure}

The nitrogen molecule is well known for exhibiting multireference character at long bond lengths and here methods based on a single reference can fail, see, e.g., Ref.~\onlinecite{doi:10.1080/00268970701332539}.  Hence this is another challenging system to test if MCCI can produce a potential energy curve with balanced error.  The MCCI potential curve for a cc-pVDZ basis with two frozen core orbitals is displayed in Fig.~\ref{fig:N2pot} and compared with FCI results from a number of sources.\cite{LarsenN2FCI2000,GwaltneyN2FCI2002,chanN2FCI2002}  We see that the general shape of the curve is recovered by MCCI when using a cut-off of $c_{\text{min}}=10^{-3}$ and the gap appears fairly constant except at very small bond lengths. For the values for which FCI results could be found, depicted in Fig.~\ref{fig:N2pot}, the NPE is $6.6$ kcal/mol, while the MCCI CSFs ranged from about a thousand to almost five thousand as the bond length increased with a mean value of $2,854$. This is in contrast to a SD space of $4.3\times10^{9}$ for a FCI when ignoring spatial symmetries.  The $200$ iterations required from around $3$ minutes at small $R$ to $1.3$ hours as $R$ became large using $8$ processors.

For the six bond lengths considered in Ref.~\onlinecite{chanN2FCI2002} for this system using the density matrix renormalization group (DMRG) the NPE in $mE_{h}$ ranged from $0.4$ to $0.01$ for DMRG1000 and DMRG4000 respectively while CCSD(T) was $172.724$ and CCSD $57.754$.  Using a CAS(6, 6) wavefunction as a reference, MRCI and MRCC had $0.464$ and 0.$732$ $mE_{h}$ respectively, but the number of states used was not specified. While the results of our MCCI calculation with $c_{\text{min}}=10^{-3}$ gave $5.37mE_{h}$ but using only a few thousand CSFs on average. The variational parameters of DMRG allow a rough comparison with the size of the space. The number of parameters is known to be $O(M^{2}k)$ (see for example Ref.~\onlinecite{ChanDMRGreview2011}) and so for the $26$ orbitals non-frozen orbitals used there are of the order of $26$ million parameters for DMRG1000 ($M=1000$).  Ref.~\onlinecite{DAS10multiCC} uses the same set of points to compare the accuracy of multireference coupled cluster approaches, again with a CAS(6, 6) reference. For state-specific multireference CCSD (SS-MRCCSD) with delocalised orbitals they find an NPE of $12.0$ $mE_{h}$, single-root MR Brillouin-Wigner CCSD (srMRBWCCSD) had an NPE of $19.4$ $mE_{h}$, and MR averaged quadratic CC MRAQCC gave $1.1$ $mE_{h}$.

   In Ref.~\onlinecite{SearsMulti05}, the thirteen points  of Ref.~\onlinecite{LarsenN2FCI2000} are used. Over these points we calculate the MCCI NPE as $5.9$ kcal/mol.  The CASSCF(10, 10) `1:1 active space' calculation\cite{SearsMulti05} had the highest NPE at $22.93$ kcal/mol of the methods they considered for this system, while the CASSCF(10, 8) `valence active space' result had an NPE of $15.03$ kcal/mol. Using their minimal active space, defined with knowledge of which bonds were broken, resulted in a CASSCF(6, 6) with a NPE of $14.59$ kcal/mol.  Interestingly the CASPT2(6, 6) had the largest NPE of the three active spaces considered for CASPT2 at $5.2$ kcal/mol, unlike their results for HF and \subscript{CH4}, while CASPT2(10, 10) had the smallest at  $1.88$ kcal/mol. The SOCI were all below one kcal/mol but it was not clear how many states were used to achieve this result. MCCI with the cut-off employed is hence relatively close in accuracy to the CASPT2 results where knowledge of which bonds are broken has been used and it appears that large active spaces for CASPT2 are needed to reduce the NPE to below $2$ kcal/mol. 
 
\begin{figure}[ht]\centering
\includegraphics[width=.45\textwidth]{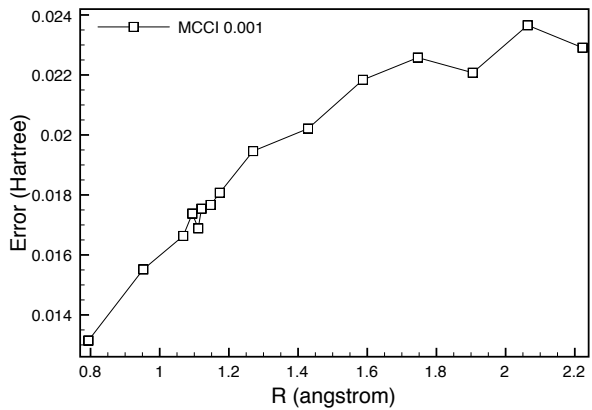}
\caption{Error (Hartree) in the MCCI ($c_{\text{min}}=10^{-3}$) calculation when compared with the FCI against bond length $R$ (angstrom) for N\subscript{2} using the cc-pVDZ basis set with two frozen cores.  The NPE is $6.6$ kcal/mol.}\label{fig:N2poterror}
\end{figure}

In Fig.~\ref{fig:N2poterror} we see that the difference between the MCCI and FCI energy tends to increase with the bond length for the nitrogen molecule.

\begin{figure}[ht]\centering
\includegraphics[width=.45\textwidth]{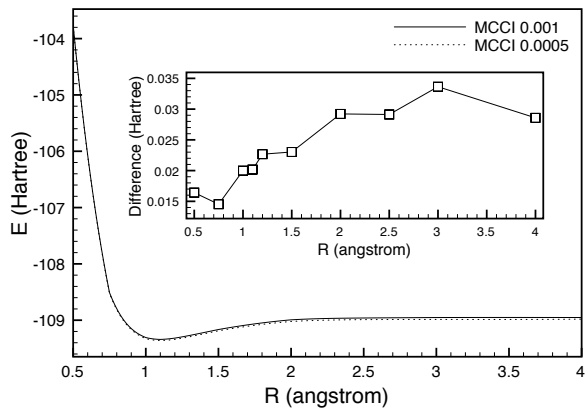}
\caption{MCCI energy (Hartree) against bond length $R$ (angstrom) for N\subscript{2} using the cc-pVTZ basis set with a cut-off of $c_{\text{min}}= 10^{-3}$ or $c_{\text{min}}=5\times 10^{-4}$. Inset: Energy difference between the MCCI results using the two cut-offs.}\label{fig:N2potVTZ}
\end{figure}

We have also calculated a potential curve using a cc-pVTZ basis. This is displayed in Fig.~\ref{fig:N2potVTZ} and shows how the dissociation is again well-behaved  for this larger range of points. Without symmetry considerations the FCI space is around $10^{17}$ SDs while the mean number of CSFs required for the MCCI curve was around $4,600$  at $c_{\text{min}}=10^{-3}$ and around $10,600$ at $c_{\text{min}}=5\times10^{-4}$ . The very large space precludes a comparative FCI calculation. However we may make an approximate comparison with FCI results\cite{N2AnoFCI} which used an ANO[4s3p1d] basis with two frozen cores. The difference in energy between the MCCI results at the equilibrium bond length of $1.098$ angstrom and at $4$ angstrom is $0.381$ Hartree at the smallest cut-off used. This compares reasonably well with the approximate dissociation energy FCI results of Ref.~\onlinecite{N2AnoFCI}.  There an approximate equilibrium bond length of $2.1$ Bohr was used to give $0.321$ Hartree and $40$ Bohr was used to calculate the energy at dissociation.   When comparing the MCCI energy of the system at $R=4$ angstrom with the sum of the MCCI energies of the appropriate fragments we find a difference of $56.8$ kcal/mol when using $c_{\text{min}}=10^{-3}$ and a difference of $41.8$ kcal/mol when using $c_{\text{min}}=5\times10^{-4}$. This suggests that the atoms are treated more accurately when using MCCI with HF MOs on this multireference system with a  basis that is beyond current FCI. However we note that the curve appears to dissociate correctly even with the largest difference seen in this work between the molecule essentially at dissociation and the fragments suggesting that MCCI is the least size-consistent in this case.

To give a somewhat fairer comparison with the FCI results of Ref.~\onlinecite{N2AnoFCI} we remove the f orbitals from a cc-pVTZ basis, and attempt to use the same geometries as Ref.~\onlinecite{N2AnoFCI}, but keep the MCCI cut-off value the same and do not freeze any core orbitals. For $R=2.1$ Bohr we find an energy of $-109.296$ Hartree using $12,664$ CSFs, but the MCCI results were much too high in energy at $R=40$ Bohr ($-93.66$ Hartree using $5,029$ CSFs) which we attribute to the restricted Hartree-Fock (RHF) reference being qualitatively wrong.  The amount of states needed to compensate for this and produce a sensible MCCI energy using the RHF molecular orbitals is then perhaps much too large for the cut-off employed. However we saw that when using a full cc-pVTZ basis the potential curve dissociated as one would expect in Fig.~\ref{fig:N2potVTZ} when the largest bond length was $4$ angstrom.   Hence we note that the problems resulting in a very poor reference at extremely large bond length could be circumvented, to a degree, by calculating the curve until approximate convergence towards the energy as dissociation where the smaller bond distance means that RHF is not quite as incorrect.  To this end we use $R=7.56$ Bohr ($4$ angstrom) and find the energy to be $-108.959$ with $19,156$ CSFs. This gives an approximate dissociation energy of $0.337$ Hartree. The MCCI result therefore has an approximate error of around $10$ kcal/mol.  The symmetry adapted FCI space was around $9.68\times 10^{9}$ while the mean number of CSFs used by MCCI for the two points was $15,910$.  Hence here only around $1.6\times 10^{-4}\%$ of the FCI symmetry adapted space was used. 

\subsection{BeH\subscript{2}}

\begin{figure}[ht]\centering
\includegraphics[width=.45\textwidth]{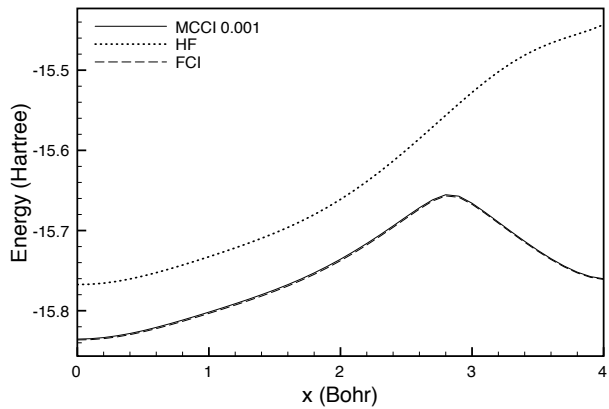}
\caption{ MCCI ($c_{\text{min}}=10^{-3}$), FCI and Hartree-Fock energy (Hartree) against the reaction coordinate $x$ (Bohr) for BeH\subscript{2} using the cc-pVDZ basis. }\label{fig:BeH2}
\end{figure}

The model reaction for the formation of BeH\subscript{2} was put forward in Ref.~\onlinecite{PurvisBeH2} to investigate CCSD. There CCSD was found to describe the system well even when the system had multireference character with two states.  Recently, BeH\subscript{2} has been considered as a test system for internally contracted multireference coupled cluster (ic-MRCC)\cite{evangelista:114102}.  In the aforementioned work the beryllium atom is at the origin and the hydrogen atoms have, in Bohr, the coordinates $x,y$ and $x,-y$ where $y=2.54-0.46x$ and $x\in[0,4]$.  We run the calculations with no frozen orbitals and note that the full CI space is around $4$ million SDs when neglecting spatial symmetries, while the MCCI wavefunction with $c_{\text{min}}=10^{-3}$ consisted of CSFs ranging from $386$ at $x=0$ to a maximum of $1,002$ around the transition state at $x=2.8$ with a mean value of $628$.  Full CI results were calculated using MOLPRO.\cite{MolproFCI1,MolproFCI2,MOLPRO}  We see that the match between the MCCI and the FCI is both qualitatively correct and quantitatively appears very good in sharp contrast to the incorrect behaviour of Hartree-Fock here (Fig.~\ref{fig:BeH2}).  The good match of the MCCI with the FCI is shown in the NPE value of $0.628$ kcal/mol despite the wavefunction only being a few-hundredths of a percent of the size of the FCI SD space.  We note that in Ref.~\onlinecite{evangelista:114102} the NPE, when using a slightly different basis set, was found to be $0.653$ kcal/mol for ic-MRCCSD.

\begin{figure}[ht]\centering
\includegraphics[width=.45\textwidth]{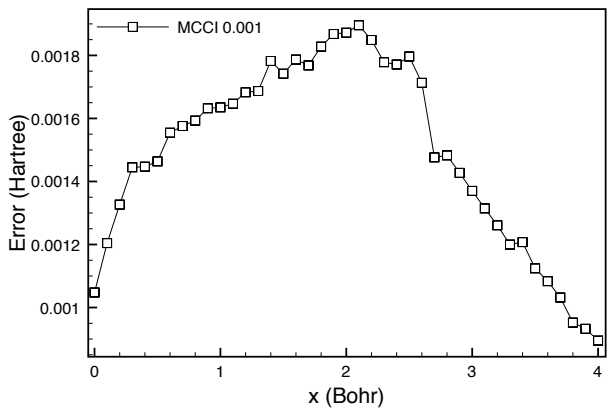}
\caption{Error (Hartree) in the MCCI ($c_{\text{min}}=10^{-3}$) energy compared with FCI against the reaction coordinate $x$ (Bohr) for BeH\subscript{2} using the cc-pVDZ basis. The NPE is $0.63$ kcal/mol.}\label{fig:BeH2error}
\end{figure}

We see in Fig.~\ref{fig:BeH2error} that the difference between the FCI and MCCI energy is smallest at $x=4$ Bohr and increases as the reactants approach ($x$ decreases) until about $x=2$ Bohr from then on it decreases as $x$ decreases.  We note that the maximum error does not correspond to the transition state.

\subsection{Ammonia inversion}

We investigate the ability of MCCI to reproduce the potential curve of NH\subscript{3} as its trigonal pyramid inverts by moving through a planar structure. We use the cc-pVDZ basis and freeze one orbital. The NH bond length is $1.025$ angstrom and the hydrogens are at $120$ degrees to each other. The NH bond makes an angle $\theta$ with a line passing through the nitrogen and perpendicular to the hydrogen plane.  

\begin{figure}[ht]\centering
\includegraphics[width=.45\textwidth]{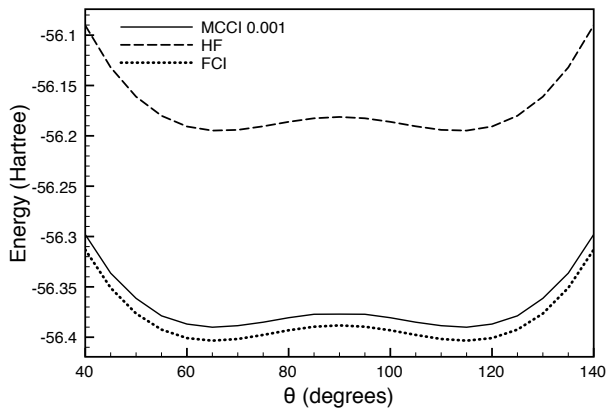}
\caption{ MCCI ($c_{\text{min}}=10^{-3}$), FCI, and Hartree-Fock energy (Hartree) against angle ($\theta$) for NH\subscript{3} using the cc-pVDZ basis and one frozen core orbital. }\label{fig:Ammmoniacurve}
\end{figure}

The MCCI curve is displayed in Fig.~\ref{fig:Ammmoniacurve} and compared with Hartree-Fock and FCI results which were calculated using MOLPRO.\cite{MolproFCI1,MolproFCI2,MOLPRO}  The plot is mirrored around the line $x=90$. We see that both MCCI with $c_{\text{min}}=10^{-3}$ and Hartree-Fock recover the shape of the FCI curve with the minima and transition point in approximately the correct place.  The MCCI curve appears to be a little too flat as it approaches the planar structure in that the energy does not change as much as the FCI or Hartree-Fock between $85$ and $90$ degrees. Despite this, for the $11$ points from $40$ to $90$ degrees, the NPE is $2.4$ kcal/mol while Hartree-Fock has an NPE of $9.9$ kcal/mol.  The number of CSFs ranged from $1,226$ to $1,824$ with a mean value of $1,629$.  This is in contrast to the size of the FCI SD space when neglecting symmetry of $\sim 4\times 10^{8}$. 

\begin{figure}[ht]\centering
\includegraphics[width=.45\textwidth]{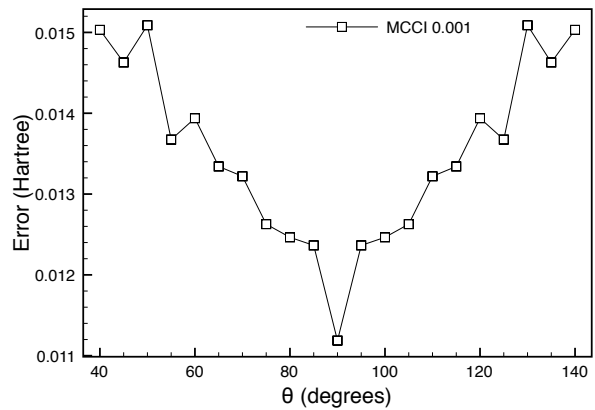}
\caption{ Error (Hartree) in the MCCI ($c_{\text{min}}=10^{-3}$)  energy when compared with the FCI against angle ($\theta$) for NH\subscript{3} using the cc-pVDZ basis and one frozen core orbital. The NPE is $2.4$ kcal/mol.}\label{fig:AmmmoniacurveError}
\end{figure}

The difference in energy between the MCCI and FCI calculation is displayed in Fig.~\ref{fig:AmmmoniacurveError}.  For this system we see that the error generally decreases as the planar transition state is approached.

\subsection{Hydrogen lattice}

We now consider a linear chain of hydrogen atoms.  Such a system would perhaps be expected to be more suitable for modelling using techniques appropriate for strongly-correlated one-dimensional lattice systems such as density-matrix renormalization group (DMRG)\cite{DMRGchemWhite99} methods which can in principle scale linearly with the size of the system;  a linear $50$ Hydrogen system has been considered using the DMRG for quantum chemistry in Ref.~\onlinecite{ChanHchain2006} with a STO-6G basis.  The 1D hydrogen chain therefore poses an interesting challenge for MCCI particularly as the correlation progresses from dynamic to a large amount of, what could be considered, static correlation as the distances between the atoms increase: the FCI results using MOLPRO\cite{MolproFCI1,MolproFCI2,MOLPRO} for a chain of $12$ hydrogens in the STO-6G basis have only two coefficients greater than $0.05$ ( $-0.110$  and $0.981$) at $1.0$ Bohr separation compared with $19$ at $4.2$ Bohr where the largest is only $0.22$. 

\begin{figure}[ht]\centering
\includegraphics[width=.45\textwidth]{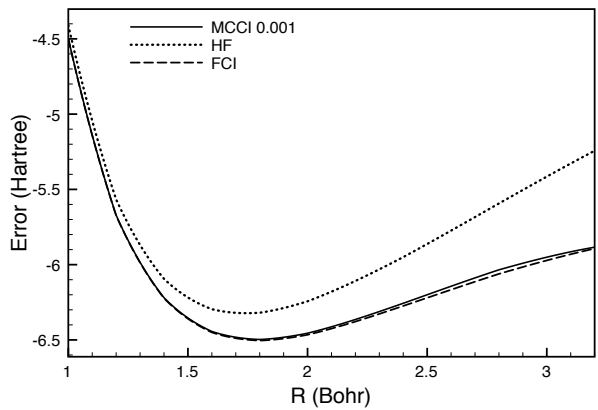}
\caption{ MCCI ($c_{\text{min}}=10^{-3}$), FCI, and Hartree-Fock energy (Hartree) against distance between atoms $R$ (Bohr) for a chain of twelve hydrogen atoms with a STO-6G basis.}\label{fig:Hchain12pot}
\end{figure}

\begin{figure}[ht]\centering
\includegraphics[width=.45\textwidth]{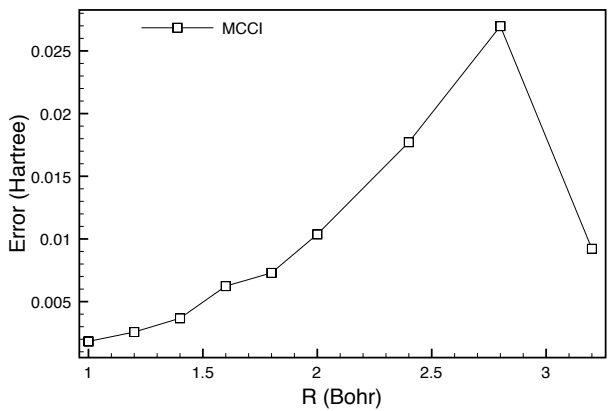}
\caption{Error (Hartree) in the MCCI ($c_{\text{min}}=10^{-3}$) energy compared with the FCI against distance between atoms $R$ (Bohr) for a chain of twelve hydrogen atoms with a STO-6G basis. The NPE is $15.8$ kcal/mol.}\label{fig:H12error}
\end{figure}

In Fig.~\ref{fig:Hchain12pot} we see that for a chain of twelve hydrogens the Hartree-Fock energy increases much too fast as the distance between hydrogen atoms increases.  MCCI, with $c_{\text{min}}=10^{-3}$, is a fairly good match with the FCI results.  We note that there was a numerical issue with the MCCI results for $R>3.2$ Bohr when using MOs.  Here MCCI produced an energy slightly lower than the FCI energy. However when using orthogonal atomic orbitals we found a MCCI energy of -5.694 Hartree at 4.2 Bohr compared with the FCI result of -5.699 Hartree, this seemed to suggest that the use of very poor MOs in MCCI for this challenging system can lead to numerical problems when using CSFs in MCCI.  The full configuration space consists of $853,776$ determinants while the MCCI wavefunction ranged from  $342$ at $R=1$ Bohr to  a maximum of $6,477$  by $R=2.8$ Bohr. The mean value was $2,701$ CSFs.  The NPE for the displayed results was $15.8$ kcal/mol. 
In Fig.~\ref{fig:H12error} we display the difference between the MCCI and FCI energies.  We see that the MCCI energy error increases from its value at $R=1$ Bohr until $R=2.8$ Bohr. 

\begin{figure}[ht]\centering
\includegraphics[width=.45\textwidth]{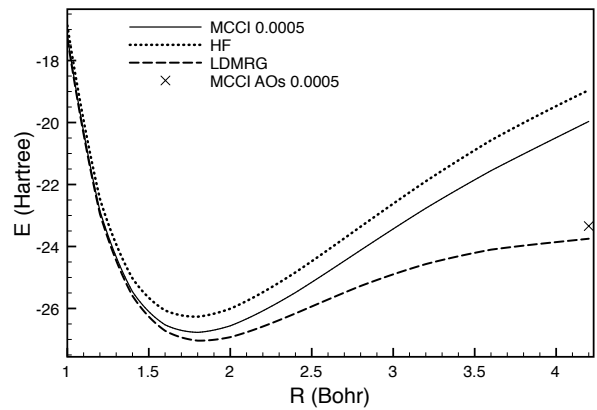}
\caption{ MCCI ($c_{\text{min}}=5\times10^{-4}$), Hartree-Fock, LDMRG(500) (see Ref.~\onlinecite{ChanHchain2006}), and MCCI ($c_{\text{min}}=5\times10^{-4}$) using orthogonal atomic orbitals (AOs). Energies (Hartree) against distance between atoms $R$ (Bohr) for a chain of fifty hydrogen atoms when using a STO-6G basis.}\label{fig:Hchain50pot}
\end{figure}

We did not satisfactorily calculate a potential curve (Fig.~\ref{fig:Hchain50pot}) for the $50$ hydrogen system compared with DMRG based results of Ref.~\onlinecite{ChanHchain2006} which we suggest as due to the extremely large configuration space ($10^{28}$) of which we only included up to $\sim 80,000$ states (circa $10^{-21} \%$ of the full space) which even assuming that the fraction of states needed for an accurate potential curve decreases with the size of the space still appears to be much too small a sample.  We note that the DMRG results used localized orbitals while the MCCI potential curve used HF MOs.  We can see in Fig.~\ref{fig:Hchain50pot} that the MCCI curve improves upon that of Hartree-Fock in this basis and retains the correct equilibrium bond length but also still displays incorrect behaviour at large $R$. This suggests that it does not sufficiently account for the static correlation for this system at the value of $c_{\text{min}}$ used ($5\times10^{-3}$).  We note that the MCCI and Hartree-Fock curve for $R \lesssim 1.8$ Bohr appear to have moderately good agreement with the DMRG results which may be expected as the correlation is essentially dynamic here. 

As $R$ becomes larger we would expect the system to be closer to a collection of non-interacting hydrogen atoms so the atomic orbitals might be expected to be a more efficient basis than the restricted Hartree-Fock molecular orbitals. We create orthogonal atomic orbitals using the Gram-Schmidt procedure starting with the leftmost atomic orbital. The use of CSFs allow us to put one electron in each atomic orbital and still have a correct $S=0$ spin function. At $R=4.2$ Bohr we find that we have an energy which is $0.4$ Hartree above the DMRG result, but the MCCI program does not seem able to easily improve upon this single CSF as no other configurations are then found.  Such an approach would not work well for slightly smaller $R$ where both dynamic and static correlation is important---for $R=3.2$ Bohr this approach yields a difference of $1.2$ Hartree---and suggests that a larger fraction of the configuration space would need to be explored and that approximate natural orbitals could be useful. 

\subsection{Ethylene torsional angle}

We now look at ethylene for which we use $R_{C-C}=1.325$ angstrom, $R_{C-H}=1.090$ angstrom and $\angle HCC=120.252$ degrees.  We use a cc-pVDZ basis with $c_{\text{min}}=10^{-3}$ and vary the torsional angle. We find the torsional barrier in ethylene to be $75.52$ kcal/mol which compares relatively favourably with the cc-pVDZ results from Ref.~\onlinecite{piris:164102}. There the barrier was found to be $68.2$ kcal/mol when using CASSCF(12,12) while CASPT2 lowered this to $65.5$ kcal/mol.  The Hartree-Fock barrier was as high as $111.8$ kcal/mol. We note that it is not clear whether exactly the same geometry was used and also whether it was kept fixed, as in our results, or made more realistic by being allowed to relax at $90$ degrees in the quoted literature values (i.e., fully geometry optimizing the transition state structure).

\begin{figure}[ht]\centering
\includegraphics[width=.45\textwidth]{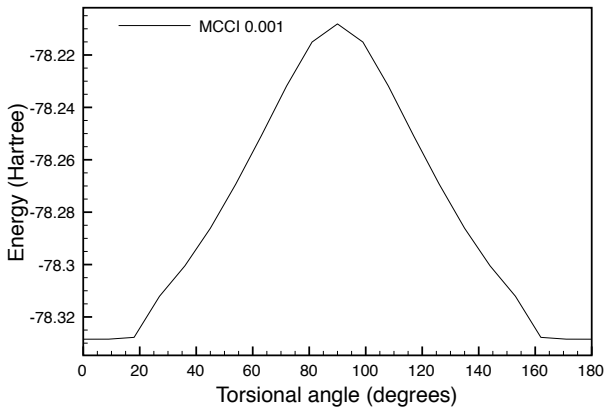}
\caption{Energy against torsional angle (degrees) for ethylene with a cc-pVDZ basis and $c_{\text{min}}=10^{-3}$.}\label{fig:Ethylene}
\end{figure}

The number of CSFs increased from around  $5,900$ to about $11,200$ as the torsional angle increased to ninety degrees. The mean number of CSFs was $8,250$ which is  $8.3\times 10^{-12} \%$ of the size of the $10^{17}$ SD FCI space when neglecting possible spatial symmetries. In Fig.~\ref{fig:Ethylene} the results have been mirrored about the line $x=90$. The lack of a cusp at $90$ degrees when using multireference perturbation but its occurrence when using single reference methods such as CCSD was noted in Ref.~\onlinecite{chaudhuri:144304} and we see here that a cusp appears to be avoided as the gradient of the MCCI curve decreases as it approaches $90$ degrees.

\section{Summary}

We have seen that potential energy curves for small systems may be calculated by MCCI to approaching chemical accuracy, when compared with full configuration interaction results, even when the system is known to be multiconfigurational. Table~\ref{tbl:AllFCIcompare} shows how NPE values of often around a few kcal/mol can be achieved using MCCI to construct a wavefunction that uses only a very small percentage of the size of the FCI space.  For all the systems but BH, which has fairly similar NPE and mean single-point energy error, and the chain of twelve hydrogens, which had the largest fraction of states but worst NPE, we achieved a better NPE than the mean single point error demonstrating that MCCI can indeed be used for calculating a potential curve to higher accuracy than its constituent single-point energies.  We note that in general for a fixed $c_{\text{min}}>0$ MCCI would not be expected to be size-consistent nor size-extensive, but the potential curves calculated here suggest, that for a low enough $c_{\text{min}}$, MCCI is sufficiently size-consistent to correctly describe the approach to dissociation of many of the systems considered in this work.  Quantifying the size-consistency by comparing the MCCI energy at the largest bond length and lowest cut-off used with the sum of the MCCI energy of the fragments reveals a difference of less than 10 kcal/mol for most of the systems considered and less than 1 kcal/mol for those undergoing a single hydrogen dissociation, however N\subscript{2} in a cc-pVTZ basis had a difference of 41.8 kcal/mol.

  We note that as the space became larger the fraction used tended to decrease, for example the results for the nitrogen and fluorine molecules in Table~\ref{tbl:AllFCIcompare}, which we would hope may be a general feature.  We acknowledge that the NPE was larger for the nitrogen and fluorine molecules when the space became larger but comparing, e.g., nitrogen with BH we see that the NPE is only around $2.5$ times larger but the fraction of the space is around $100$ times smaller while the reduction in the space is even more pronounced for fluorine and the increase in NPE a little smaller.  We note that the NH\subscript{3} FCI has more than six times the states of BH but MCCI uses a fraction around $100$ times smaller to give a slightly lower NPE.  We also saw that nitrogen in a cc-pVDZ basis used $2,854$ on average of a possible $~10^{9}$ states while in a cc-pVTZ basis without frozen orbitals we could produce a well-behaved curve using just $10,600$ states, at the smallest cut-off considered, of the much larger $10^{17}$ FCI space, although an NPE value was not available in the latter case. We acknowledge that a fixed value of $c_{\text{min}}$ restricts the size of the MCCI space so that the maximum fraction possible must decrease as the FCI space enlarges.  However our MCCI results are well within this limit and we note that the number of MCCI states appears to increase much more slowly than the FCI space.
\begin{table}[h]
\centering
\caption{Table showing the  mean CSFs to FCI SDs (without symmetry considerations) ratio, mean CSFs to FCI CSFs (without symmetry considerations),  Non parallelity error (NPE) (kcal/mol) and Mean single-point error (MSPE) (kcal/mol), at the smallest cut-off used, for the systems investigated for which FCI results were available. } \label{tbl:AllFCIcompare}
\begin{tabular*}{9.5cm}{@{\extracolsep{\fill}}lcccc}
\hline
\hline
System & $\%$ FCI SD space & $\%$ FCI CSF space & NPE & MSPE    \\
\hline
  BH & 0.007$\%$ & 0.021$\%$ &2.6 & 2.1 \\
  HF & 0.014$\%$ & 0.056$\%$     &1.3  & 2.6\\
   CH\subscript{4} & 0.012$\%$ &  0.049$\%$   &0.6 & 4.1 \\
   C\subscript{2} & 0.003$\%$  &  0.013$\%$   &4.9 & 6.0\\
   F\subscript{2} & $8.3\times 10^{-7} \%$ & $4.9\times10^{-6}\%$  & 6.2 &  9.0   \\
   N\subscript{2} & $6.6 \times 10^{-5} \%$    &  $3.2\times10^{-4}\%$ & 6.6 & 11.9\\
   H\subscript{12}& 0.32$\%$ & 1.2$\%$   & 15.8 & 6.0\\
   BeH\subscript{2}  & 0.016 $\%$   & 0.052$\%$    &0.63 & 0.94\\
   NH\subscript{3} &  $4\times 10^{-4}\%$ &  $1.7\times10^{-3}\%$   &2.4 & 8.4\\   
\hline
\hline
\end{tabular*}

\end{table}

  However, even with the possibility of a decrease in the fraction of the FCI space required as the system became very much larger, we found that a chain of fifty hydrogens was to prove too challenging for MCCI.  Here we saw that although the equilibrium geometry was correct this was not an improvement on the Hartree-Fock results and for longer bond lengths, where static correlation would be considered important, the curve, although lower in energy than that of Hartree-Fock, increased too quickly.  We suggest that this is due to the very large size of the configuration space $(\sim 10^{29})$ and the very challenging system for standard quantum chemistry methods partly due to its strongly correlated nature: restricted Hartree-Fock is a very poor approximation at large bond lengths here and H\subscript{12} required the largest fraction of the FCI space and gave the largest NPE. We note that as a single CSF using orthogonalised atomic orbitals was much closer to the DMRG energy here then the results could perhaps be improved by the use of approximate natural orbitals and allowing the algorithm to explore more of this space but this strongly correlated system may still require many states and seems to be more amenable to modelling using other methods. Finally, the potential curve for the isomerisation of ethylene was found to be relatively smooth and the calculated barrier compared fairly favourably with other computational results. 
Although we calculated potential surfaces, for which FCI results exist, to relatively high accuracy using just a very small fraction of the FCI space, we have noted two possible limitations of MCCI when applied to potential energy surfaces. The first is that when the cut-off is not sufficiently small a potential curve may not be smooth due to the stochastic nature of the algorithm.  However, we were able to produce smooth curves with reasonable cut-offs that resulted in only a very small fraction of the FCI space being used. The second limitation is when many states of a large FCI space are required due to a combination of the reference being very qualitatively wrong and the system being considered strongly correlated, e.g., in the case of restricted Hartree-Fock used with N\subscript{2} at extremely large bond lengths or H\subscript{50} at moderately large bond lengths, then not enough states at a feasible cut-off can be included to compensate for the poor choice of molecular orbitals.  For N\subscript{2} it seemed that this could be remedied to some degree by considering a smaller bond length where the system had almost converged to the energy at dissociation. However this was not possible for H\subscript{50} so we suggested that different orbitals, perhaps approximate natural orbitals, might be used to try and improve the description at bond lengths away from the equilibrium geometry.  To have a better chance of working well with extremely large FCI spaces we also suggest that MCCI could perhaps be used with a relatively large cut-off to find a good starting point for multireference perturbation theory.

\acknowledgements{We thank the European Research Council (ERC) for funding under the European Union's Seventh Framework Programme (FP7/2007-2013)/ERC Grant No. 258990.}

\providecommand{\noopsort}[1]{}\providecommand{\singleletter}[1]{#1}%

\end{document}